\documentstyle[aps,preprint]{revtex}
\begin{document}
\draft
\hsize\textwidth\columnwidth\hsize\csname @twocolumnfalse\endcsname

\title{
Exact Ground States of One-Dimensional Quantum Systems: Matrix Product Approach
}
\author{Gang Su$^\ast$
}
\address{
 Institut f\"ur Theoretische Physik,
Universit\"at zu K\"oln\\
 Z\"ulpicher Strasse 77, D-50937 K\"oln, Germany\\
}
\maketitle

\begin{abstract}
By using the 
so-called matrix-product ground state approach, a few one-dimensional quantum
systems, including a frustrated spin-1/2 
Heisenberg ladder, the ferromagnetic t-J-V model at half-filling, the antiferromagnetic $J_z-V$ at 
2/3 filling and the antiferromagnetic $t-J_z-V$ model at half-filling,
 are solved exactly. The correlation functions in
the ground states are calculated respectively. Some relevant results are also
discussed.

\end{abstract}


\section{Introduction}

Recently there has been growing interest on one-dimensional (1D) quantum systems 
due to a variety of reasons. Since the difficulty in dealing with many-body
problems the methods in obtaining exact results in 1D are rare. As is 
well-known, a few 1D many-body problems can be exactly solved by the Bethe 
ansatz, as the systems satisfy the so-called Yang-Baxter equation\cite{baxter}. Actually,
Bethe ansatz is a powerful and efficient method in obtaining exact results
including ground state and excited states as well as thermodynamics. However,
a lot of 1D quantum systems does not obey Yang-Baxter equation, and thus are
non-integrable. In this situation, it is difficult, due to absence of 
a systematic method, to obtain some exact results of 1D many-body systems.
Amongst very few existing exact 
methods
there is a so-called matrix product (MP) ground state approach which can be 
applied to obtain 
exact ground-state properties of
1D quantum systems. This method was established a few 
years ago by Kl\"umper et al\cite{klu} where they studied the exact ground 
state
of a large class of antiferromagnetic spin-1 chains. In those cases the method
was really efficient. The basic idea of the method is that the global ground 
state can be constructed in the form of a matrix product of single-site states.
This state is usually said optimal in the sense that it is the product of 
all the
local ground states of the local interaction $h_{j,j+1}$ (see below). To show
that the MP state is really the ground state, one can diagonalize $h_{j,j+1}$
exactly in all local eigenstates. If the eigenvalue of $h_{j,j+1}$ in the MP 
ground state is not larger than
the minimum of all eigenvalues of $h_{j,j+1}$ 
diagonalized in all local eigenstates, then the MP state is the global ground 
state of the whole Hamiltonian. The similar spirit has been applied to 
construct the exact ground states of generalized Hubbard models\cite{boer}. In
this paper, we will apply the MP approach to a few quantum systems including
a frustrated spin-1/2 Heisenberg spin ladder, and several simplified cases of
t-J-V model. The correlation functions of the ground states are obtained. Some
relevant results are also addressed.   

\section{A Frustrated Spin-1/2 Heisenberg Ladder}

Investigations on spin ladders have attracted a lot of attention in recent
years\cite{Dago1,Dago,Strong,Barnes,Rice,Troy,White,Shel,xian}. This is usually 
motivated from the following possible 
reasons: one is that the Heisenberg spin ladder (two antiferromagnetically 
coupled Heisenberg antiferromagnetic spin-1/2 chains or
two ferromagnetically coupled such chains\cite{Taka,Wata}) provides a simple model to gain insight
into the underlying physics of crossover from 1D chain to 2D
square lattice; one is due to the Haldane's conjecture\cite{Hald} $-$ which states that a
quantum antiferromagnet with integral spin has different properties from one
with half-integral spin, i.e., the former has a unique ground state, a gap
to the excited state and exponential decaying ground-state correlation 
functions, which have been confirmed in experiments, 
while the latter does not $-$ people believe that the Heisenberg spin ladder
is probably in the same phase as the Heisenberg spin-1 antiferromagnetic chain\cite{Troy,White,xian};
 and the other is that spin ladders are intimately related to magnetic properties of some
realistic materials (for example, $Sr_{n-1}Cu_{n+1}O_{2n}$\cite{Takano,Rice}
and $(VO)_2P_2O_7$\cite{Eccle}). Besides, frustrated spin ladders are also
interesting, because they are
believed to be responsible for physical properties of stoichiometric 
$Sr_{n-1}Cu_{n+1}O_{2n}$ compounds\cite{Rice}. 
Exact solutions on spin ladders are rare so far, 
whereas earlier studies on these models are either approximate 
(mean-field treatment or bosonization) or numerical, although some consensus 
is qualitatively made. Any exact result
on these models will hence be necessary.
In this section, we report that a frustrated spin-1/2 Heisenberg ladder, with
properly constricted parameters, can be solved exactly by using the 
MP ground state approach. 

We consider a spin-1/2 Heisenberg ladder with diagonal coupling, 
described by the following Hamiltonian

\begin{eqnarray}
&& H = J_1 \sum_{n=1,2} \sum_{j=1}^{L} {\bf S}_{n}(j) \cdot {\bf S}_{n}(j+1)
+ J_2 \sum_{j=1}^{L}{\bf S}_{1}(j) \cdot {\bf S}_{2}(j) \nonumber \\ && \hspace{0.5cm}
+ J_3 \sum_{j=1}^{L} [{\bf S}_{1}(j) \cdot {\bf S}_{2}(j+1) 
 + {\bf S}_{2}(j) \cdot {\bf S}_{1}(j+1)] \nonumber \\ && \hspace{0.5cm}
+ \frac{2(J_1 + J_3) + J_2}{4}, 
\end{eqnarray}
where $j$ runs over all rungs and $n$ over two legs: 1 and 2, $L$ is the length
of the ladder, ${\bf S}_{n}(j)$ is spin-1/2 operator on the rung 
$j$ along leg $n$ of the ladder, $J_1$ is the interaction along the legs,
$J_2$ is the coupling along the rungs, and $J_3$ is the diagonal coupling 
between the two nearest neighbor rungs. A constant (the last term of (1)) is
added for convenience of our purpose. The periodic boundary condition is assumed along the ladder.
To guarantee the solvability of the 
ladder, we have to confine the parameters to satisfy the following conditions:
\begin{eqnarray}
0 < J_1 \leq - \frac{J_2J_3}{J_2 + 2 J_3}, ~~J_2 < 0, ~~ J_3 < 0.
\end{eqnarray}
Under the conditions of (2), one may see that spins on the ladder are
frustratedly distributed along the legs and rungs, minimizing the total energy of the system. We 
mean the frustration in the present model just in this sense. In the following we for brevity denote the local
eigenstates $|+\frac12\rangle_{j}$ and $|-\frac12\rangle_{j}$ of $S^{z}(j)$ 
simply by $|+\rangle_{j}$ and $|-\rangle_{j}$, respectively. 
We denote a local state on each rung
by $|a\rangle_{j}^{m}|b\rangle_{j}^{n}$, where $\{a,b\}=\{+,-\}$, and $\{m,n\}
=\{1,2\}$ identifying the legs. Introduce a
permutation operator
\begin{eqnarray}
P_{j,l}^{(m,n)} = \frac{1}{2}[1 + 4 {\bf S}_{m}(j) \cdot {\bf S}_{n}(l)].
\end{eqnarray}
The operator has the property:
\begin{eqnarray}
P_{j,l}^{(m,n)}|a\rangle_{j}^{m}|b\rangle_{l}^{n} = 
|b\rangle_{j}^{m}|a\rangle_{l}^{n}.
\end{eqnarray}
With the aid of (3), we rewrite (1) as
\begin{eqnarray}
H = \sum_{j=1}^{L}h_{j,j+1},
\end{eqnarray}
with the local interaction
\begin{eqnarray}
&& h_{j,j+1} = \frac{J_1}{2}(P_{j,j+1}^{(1,1)} +  P_{j,j+1}^{(2,2)}) + 
\frac{J_2}{4}(P_{j,j}^{(1,2)} + P_{j+1,j+1}^{(1,2)}) 
+ \frac{J_3}{2} (P_{j,j+1}^{(1,2)} +  P_{j,j+1}^{(2,1)}),
\end{eqnarray}
where we have made use of the periodic boundary condition along the ladder.

Following the discussions in Ref.\cite{klu}, and using the local eigenstates $|+\rangle_{j}$ and $|-\rangle_{j}$ of 
$S_{j}^z$,
we define a local $2\times 2$ matrix on each rung by
\begin{eqnarray}
g_{j} = \left( \begin{array}{cc}
         |+\rangle_{j}^{1}|-\rangle_{j}^{2} + |-\rangle_{j}^{1}|+\rangle_{j}^{2} & |+\rangle_{j}^{1}|+\rangle_{j}^{2} + |-\rangle_{j}^{1}|-\rangle_{j}^{2} \\
|+\rangle_{j}^{1}|+\rangle_{j}^{2} + |-\rangle_{j}^{1}|-\rangle_{j}^{2} & 
|+\rangle_{j}^{1}|-\rangle_{j}^{2} + |-\rangle_{j}^{1}|+\rangle_{j}^{2} 
                \end{array} \right)
\end{eqnarray}
and a global state
\begin{eqnarray}
|\Psi_0\rangle = Tr g_1 \otimes g_2 \otimes \cdots  \otimes g_L
\end{eqnarray}
where $\otimes$ denotes usual matrix multiplication of $2\times 2$ matrices 
with a tensor product of the matrix elements. It can be found that
\begin{eqnarray}
H |\Psi_0\rangle = E_0 |\Psi_0\rangle, \\
E_0  = (J_1 + \frac{J_2}{2} + J_3)L.
\end{eqnarray}
As we have the constraint (2), thus $E_0 < 0$. Now let us show $|\Psi_0\rangle$
is a ground state of $H$. To realize this purpose, we in turn need to prove $g_j \otimes
g_{j+1}$ and thus $|\Psi_0\rangle$ to be the ground state of $h_{j,j+1}$. After a little tedious 
algebraic calculations, one may find that 
$$h_{j,j+1} (g_j \otimes g_{j+1}) = (J_1 + \frac{J_2}{2} + J_3) (g_j \otimes g_{j+1}),$$
namely, $g_j \otimes g_{j+1}$ is an eigenstate of $h_{j,j+1}$ with eigenvalue
$E_0/L$. On the other hand, $h_{j,j+1}$ can be exactly diagonalized in all possible
16 eigenstates on a plaquette consisting of two rungs ($j$ and $j+1$) and 
two corresponding legs. It is not hard to obtain the eigenvalues as follows: 
$J_1$ (3-fold),
 $\frac{J_2}{2}$ (3-fold), $J_3$ (3-fold), $J_1 + \frac{J_2}{2} + J_3 $ (5-fold),  $
\pm [J_1^2 -\frac12 J_1J_2 + \frac14 J_2^2 - J_1J_3 - \frac12 J_2J_3 + J_3^2]^{\frac12}$.
One can easily check that $E_0/L =J_1 + \frac{J_2}{2} + J_3$ is the lowest 
eigenvalue between them, under the conditions of (2). Thus, 
$g_j \otimes g_{j+1}$ is really the
ground state of $h_{j,j+1}$, and therefore $|\Psi_0\rangle$ is also the ground state
of $h_{j,j+1}$, which in turn proves that $|\Psi_0\rangle$ is the
global ground state of $H$, and $E_0$ is the ground-state energy. 

Similar to the calculations in Ref.\cite{klu}, we can obtain 
correlation functions in $|\Psi_0\rangle$ state by using the transfer matrix 
method. The norm $\langle\Psi_0|\Psi_0\rangle$ can be obtained by
\begin{eqnarray}
\langle\Psi_0|\Psi_0\rangle = \sum_{\{n_{\alpha}, m_{\alpha}\}}g_{n_1n_2}^{+}g_{n_2n_3}^{+} \cdots g_{n_Ln_1}^{+} 
 g_{m_1m_2}g_{m_2m_3} \cdots g_{m_Lm_1} = Tr G^{L},
\end{eqnarray}
where $G$ is the transfer matrix, with elements $G_{\mu_1 \mu_2} \equiv G_{(n_1m_1),(n_2m_2)} = g_{n_1n_2}^{+}g_{m_1m_2}$ where the ordering of 
multi-indices is chosen as usual: $\mu = 1,2,3,4$ $\leftrightarrow$ $(11), 
(12), (21), (22)$. The $4\times 4$ transfer matrix has four eigenvalues: $0$
(2-fold) and $4$ (2-fold). Thus, one can write 
$G|e_{n}\rangle = \lambda_{n}|e_{n}\rangle$,
with $\lambda_{n}$ $(n=1,2,3,4)$ the eigenvalues ($\lambda_{1,2}=0$, 
$\lambda_{3,4}=4$), and $|e_{n}\rangle$ the corresponding normalized 
eigenvectors which have alternative forms
\begin{eqnarray}
|e_{1}\rangle = \left( \begin{array}{cc}
         \frac12 \\
          \frac12  \\
           -\frac12 \\
         - \frac12
                  \end{array} \right), \  \  \ 
|e_{2}\rangle = \left( \begin{array}{cc}
         -\frac12 \\
          \frac12  \\
           -\frac12 \\
         \frac12
                  \end{array} \right), \  \  \
|e_{3}\rangle = \left( \begin{array}{cc}
          \frac12   \\
           \frac12 \\
           \frac12 \\
          \frac12
                  \end{array} \right), \  \  \
|e_{4}\rangle = \left( \begin{array}{cc}
          -\frac12   \\
           \frac12 \\
           \frac12 \\
         - \frac12
                  \end{array} \right).
\end{eqnarray}
One may verify that $\{|e_{n}\rangle\}$ forms a complete set, which can be 
used to calculate the trace in the following. Evidently, 
$\langle\Psi_0|\Psi_0\rangle = 2 \cdot 4^{L}$. The one-site expectations can
be evaluated by
\begin{eqnarray}
\langle A \rangle = \frac{\langle\Psi_0|A|\Psi_0\rangle }
{\langle\Psi_0|\Psi_0\rangle} = (Tr G^L)^{-1} Tr Z(A) G^{L-1},
\end{eqnarray}
with $Z(A)_{\mu_1\mu_2} \equiv g_{n_1n_2}^{+}Ag_{m_1m_2}$. The 2-site 
correlations of operators $A(1)$ at site $1$ and $B(r)$ at site $r$ can be 
calculated by
\begin{eqnarray}
\langle A(1)B(r) \rangle = (Tr G^L)^{-1} Tr Z(A) G^{r-2} Z(B) G^{L-r}.
\end{eqnarray}
In accordance with (12)-(14), we can obtain the following expectations:
\begin{eqnarray}
 && \langle S_{1,2}^{z}(j) \rangle = 0, \  \
\langle S_{tot}^{z} \rangle = \sum_{j=1}^{L}\langle S_{1}^{z}(j) + 
S_{2}^{z}(j)\rangle = 0, \  \ \langle S_{1,2}^{+}(j)\rangle = 0,\\ &&
\langle S_{1,2}^{z}(1)S_{1,2}^{z}(r)\rangle = 
\langle S_{1,2}^{z}(1)S_{2,1}^{z}(r)\rangle  = 0, \\ &&
\langle (S_{1,2}^{z}(j))^{2}\rangle = \langle S_{1,2}^{+}(1)S_{1,2}^{-}(r)\rangle = 
\langle S_{1,2}^{+}(1)S_{2,1}^{-}(r)\rangle = \frac14,
\end{eqnarray}
for $r \geq 2$, where the technical details refer to Ref.\cite{klu}. 
We observe that in the ground state the model has
transverse fluctuations for the transverse correlation functions
are uniform with varying spatial distances, implying the correlation length
for transverse magnetic orderings is infinite, while it does not exhibit 
longitudinal magnetic order and fluctuations. As can be seen, 
$|\Psi_0\rangle$ is XY magnetic ordered. 

To identify the structure of the state, one can use a matrix 
$ u = \left( \begin{array}{cc}
         1 & 1\\
         1 & -1 
       \end{array} \right)
$ to make a similarity transformation on $g_j$ as $g_j' = ug_ju^{-1} = \left( \begin{array}{cc}
         a+b & 0\\
         0 & a-b 
       \end{array} \right)
$ with $a = |+\rangle_{j}^{1}|-\rangle_{j}^{2} + |-\rangle_{j}^{1}|+\rangle_{j}^{2}$ and $b =|+\rangle_{j}^{1}|+\rangle_{j}^{2} + |-\rangle_{j}^{1}|-\rangle_{j}^{2}$. Thus $|\Psi_0\rangle = Tr g_1' \otimes g_2'\otimes \cdots\otimes 
g_L' = |a+b\rangle_{1}|a+b\rangle_{2} \cdots |a+b\rangle_{L} + |a-b\rangle_{1}|a-b\rangle_{2} \cdots |a-b\rangle_{L}$. Obviously,
the global state is a linear combination of a direct product of all possible 
spin configurations on each rung, and thus has transverse fluctuations but 
without longitudinal ones, as revealed by the expectations. In addition, since 
$g_j$ is a mixture of singlet and triplet states, the excitation from the
ground state $|\Psi_0\rangle$ may be gapless. 

A simple argument can show that the ground state is degenerate. For instance,
if one replaces the four entries of matrix $g_j$ in (7) by the same expression:
$|+\rangle_{j}^{1}|+\rangle_{j}^{2}$ or
$|-\rangle_{j}^{1}|-\rangle_{j}^{2}$, one
will find that the replaced MPG state has the same eigenvalue as that of 
$|\Psi_0\rangle$. Actually, such states are nothing but the fully polarized 
ferromagnetic states, which can be implemented through a similarity matrix
transformation on $g_j$. One can also compute expectations of some physical
quantities in the polarized ferromagnetic state, and will find that they have
different values as in $|\Psi_0\rangle$ state, while the longitudinal 
correlation functions are still uniform in spatial space. 
Hence $|\Psi_0\rangle$ and the fully polarized ferromagnetic state are not the
same, suggesting that the ground state is not unique.   

That the model has the fully polarized ferromagnetic ground state is  
not surprising, because the 
coupling along the rungs ($J_2<0$) and the diagonal coupling between the rungs ($J_3<0$) favour to form triplets and thus prefer to form ferromagnetic order,
while the strength along the legs, though it is positive and favours to form antiferromagnetic order, is smaller than the other two interactions. The 
competing result would lead to favour ferromagnetism. The present result just 
confirms the intuitive argument. However, the XY magnetic ordered ground 
state exist in
the model is yet plausible in physics. Though the spins on each rung of the
ladder prefer to form triplets, minimizing the energy of the system, the
alignment of the spins along positive direction has the equal weight as that 
along negative direction, and thus the resultant state may have transverse
magnetic fluctuations but without longitudinal magnetic order, as
can be seen clearly from the explicit form of $|\Psi_0\rangle$. 
As the ground state is
degenerate, one may anticipate that the spin rotational symmetry is 
spontaneously broken in the ground state, as it should be. To this end,
one may note that the advantage of MP ground state approach applying to this model
lies in having found a ground state with transverse magnetic fluctuations,
thereby showing the ground state of the model is not unique, and the excitation
is gapless.

To conclude this section, the frustrated spin-1/2 Heisenberg ladder, defined in (1), 
is solved exactly by using the MP ground state approach. It is shown that the ground 
state of the model is degenerate, as a magnetic ordered state with transverse 
fluctuations and a fully polarized
ferromagnetic state coexist in the ground state. The excitation from the
ground state may be gapless. The correlation functions are uniform in spatial
space, implying that the correlation length is infinite, and thus the
system is critical in the ground state. 
One may observe that this model thus provides an example that 
quasi one-dimensional frustrated Heisenberg model can exhibit magnetic order
in the ground state, although the Mermin-Wagner thermodynamic fluctuations\cite{mermin} enable the system probably not to possess such an order at finite
temperatures. We expect that the present result, on the one hand, may
shed some light on the theoretical basis of some quasi one-dimensional 
ferromagnets, and on the other hand, may provide a clue to experimentalists
to synthesize new quasi one-dimensional ferromagnetic materials, particularly
organic magnets.

\section{t-J-V model}

This model, as an extension of the usual t-J model, first proposed by 
Schlottman\cite{sltm1} for heavy-fermions, and later studied numerically by
others\cite{kiv,dago2,troy2}. Apart from the supersymmetric points\cite{sltm1}, this
model is non-integrable. The system is described by the following Hamiltonian
\begin{eqnarray}
H_{tJV} = \sum_{j=1}^{L}h_{j,j+1},
\end{eqnarray}
with the local interaction
\begin{eqnarray}
h_{j,j+1} = - t \sum_{\sigma}(c_{j,\sigma}^{\dagger}c_{j+1,\sigma} + h.c.)
+ \frac{J}{2} (S_j^+S_{j+1}^- + S_j^-S_{j+1}^+) + J_z S_j^zS_{j+1}^z
+ V n_jn_{j+1}, 
\end{eqnarray}
on 1D chain with the length $L$ even, where $c_{j,\sigma}$ is the 
annihilation operator for an electron with spin $\sigma$ 
($=\uparrow, \downarrow$) at site $j$, $\{S_{j}\}$ are spin-1/2 operators,
$n_j=\sum_{\sigma}c_{j,\sigma}^{\dagger}c_{j,\sigma}$ is the electron number
operator at site $j$, $t$ is the hopping matrix element, 
$J, J_z$ are anisotropic spin-exchange interactions, and
$V$ is the strength of density-density interaction. 
The periodic boundary conditions
are assumed.
We further suppose that, 
as usual, the double-occupancy of every site is forbidden due to the existence
of a large on-site Coulomb repulsion, i.e., each site has either one electron
(with spin up or down) or empty. Therefore, there are three possible 
electronic states at a given site $j$: $|0\rangle$, the Fock vacuum satisfying 
$c_{j,\sigma}|0\rangle \equiv 0$; $|\uparrow\rangle \equiv 
c_{j,\uparrow}^{\dagger}|0\rangle$; and $|\downarrow\rangle \equiv 
c_{j,\downarrow}^{\dagger}|0\rangle$. Since there exists such a property, the 
problem, as first noted by Schlottman\cite{sltm1} in the t-J-V model, can be 
rewritten 
in terms of spin operators corresponding to $S=1$ in the restricted Hilbert 
space, which implies that the model should have similar behaviors as that of
$S=1$ spin chain. 
This defines the t-J-V 
model. Since it is non-integrable off supersymmetric points, it is difficult
to obtain the exact results of this model. In the following we will concentrate
on a few special cases, which can be solved exactly using MP approach.

\subsection{Ferromagnetic t-J-V model at half-filling}

In this case we assume that $J_z=J<0$ and $t\geq 0$. Similar to Sec.II, we 
define a $2\times 2$ matrix 
\begin{eqnarray}
g_{j}' = \left( \begin{array}{cc}
         |\uparrow\rangle + |\downarrow\rangle & \frac{1}{\sqrt{a}}(|\uparrow\rangle - |\downarrow\rangle ) \\
\frac{1}{\sqrt{a}}(|\downarrow\rangle - |\uparrow\rangle )    & 
|\downarrow\rangle + |\uparrow\rangle 
                \end{array} \right)_j
\end{eqnarray}
with $a>0$ a parameter, and propose an ansatz for the global ground state
\begin{eqnarray}
|\Phi_1\rangle = Tr g_1' \otimes g_2' \otimes \cdots  \otimes g_L',
\end{eqnarray}
where the notation is the same as in Sec.II. Following exactly the analyses in 
Sec.II, 
it is readily show that 
$|\Phi_1\rangle$ is the global ground state, with eigenvalue $(V + \frac{J}{4})L$,
 of the ferromagnetic t-J-V model
under the following constraint:
\begin{eqnarray}
t \geq 0, ~~~J<0, ~~~V \leq -t -\frac{J}{4}.
\end{eqnarray}
The expectation values in the ground state can be calculated directly using
the transfer matrix method. Though the finite-size results are available, we are
only interested in those in the thermodynamic limit. Henceafter we mean the
result just for $L \rightarrow \infty$. It can be found that
\begin{eqnarray}
\langle S_j^z \rangle =0, ~~ \langle S_{tot}^z \rangle = \langle \sum_{j=1}^L S_{j}^z \rangle =0, ~~ \langle S_j^{\pm} \rangle = \frac12 \frac{a-1}{a+1},~~
\langle n_j \rangle =1,~~ \langle (S_j^z)^2 \rangle =\frac14,~~\langle J_j \rangle =0,
\end{eqnarray}
with $J_j = it \sum_{\sigma}(c_{j+1,\sigma}^{\dagger}c_{j,\sigma} - 
c_{j,\sigma}^{\dagger}c_{j+1,\sigma})$, the current density. 
Therefore, we see that the ground state has XY ordering and is insulating. 
The two-point correlation 
functions are evaluated
\begin{eqnarray}
\langle S_1^zS_r^z \rangle =0, ~~\langle S_1^+S_r^- \rangle = \langle S_1^-S_r^+ \rangle = \frac14, ~~\langle n_1n_r \rangle =1, 
~~ \langle n_1S_r^z \rangle =0.
\end{eqnarray}
When $a=1$, $\langle S_j^{\pm} \rangle = 0$, implying that there is a phase 
transition from one XY
ordered state with nonvanishing order parameter into another XY ordered state 
in which the order parameter vanishes while the transverse correlation function
is uniform. Since the correlation length of transverse correlation functions
is infinite, the system is critical in the ground state.
An interesting observation is that, the state is still the ground
state, if $J=0$ so long as $V\leq -t$. This fact suggests that the 
density-density interaction seems to play a significant role in the mechanism
of the origin of ferromagnetism. Similar to arguments in Sec.II, the ground 
state of the system may be degenerate with the fully polarized ferromagnetic 
state.
Since the symmetry breaking requires the degeneracy of the ground states, the
present result is reasonable, as (23) shows that the spin rotational symmetry 
is spontaneously broken in the ground state. If one makes a particle-hole 
transformation in the system, then he will find the transformed system will 
exhibit off-diagonal long-range order in the ground state.

\subsection{Antiferromagnetic $J_z-V$ model at 2/3 filling}

Define a local $2 \times 2$ matrix
\begin{eqnarray}
g_{j}'' = \left( \begin{array}{cc}
         |\uparrow\rangle + |\downarrow\rangle + |0\rangle & |\uparrow\rangle - |\downarrow\rangle - |0\rangle  \\
|\downarrow\rangle - |\uparrow\rangle  + |0\rangle    & 
- |\downarrow\rangle - |\uparrow\rangle - |0\rangle 
                \end{array} \right)_j
\end{eqnarray}
and an ansatz for the global ground state of $H_{tJV}$
\begin{eqnarray}
|\Phi_2\rangle = Tr g_1'' \otimes g_2'' \otimes \cdots  \otimes g_L''.
\end{eqnarray}
To make $|\Phi_2\rangle$ be the global ground state of $H_{tJV}$, tracing the 
exact way in Sec.II, the following 
conditions must be satisfied:
\begin{eqnarray}
t=J=0, ~~~V=\frac{J_z}{4}, ~~~J_z >0.
\end{eqnarray}
One can see that, $|\Phi_2\rangle$ is actually the global ground state, with
eigenvalue $0$, of antiferromagnetic
Ising model with density-density interaction described by the following
Hamiltonian
\begin{eqnarray}
H_{J_zV} = J_z \sum_{j=1}^{L}S_j^zS_{j+1}^z + \frac{J_z}{4}\sum_{j=1}^{L}
n_jn_{j+1}.
\end{eqnarray}
It is well-known that this model without the density-density interaction has 
been solved by Ising\cite{ising} seventy years ago. Now we have applied matrix product
to construct the global ground state of the Ising model with density-density
interaction. Applying the similar arguments in Ref.\cite{klu} one can show 
that the ground state is unique. One will see later that the state is at 2/3 
filling. The 
state is thus nontrivial and is optimal. 

The expectation values in the ground state can be obtained:
\begin{eqnarray}
\langle S_j^z \rangle =0, ~~ \langle S_{tot}^z \rangle = \langle \sum_{j=1}^L S_{j}^z \rangle =0, ~~ \langle S_j^{\pm} \rangle = 0,~~
\langle n_j \rangle =\frac23,~~ \langle (S_j^z)^2 \rangle =\frac16.
\end{eqnarray}
The spin-spin correlation function is nontrivial 
\begin{eqnarray}
\langle S_1^zS_r^z \rangle = 0.0331456 - 0.0625(-1)^r, ~~(r \geq 2)
\end{eqnarray}
which means that the global ground state is antiferromagnetic, 
and has long-range order. 
Since the system is 
localized, the ground state should be insulating.
The spin-charge 
correlation function is
\begin{eqnarray}
\langle S_1^zn_r \rangle = 0.0331456 - 0.04167(-1)^r, ~~(r \geq 2)
\end{eqnarray}
which suggests that the spin and charge are not separated in the ground state.
The other correlation functions are:
\begin{eqnarray}
\langle S_1^+S_r^- \rangle = \langle S_1^-S_r^+ \rangle = 0, ~~\langle n_1n_r
\rangle = 0.5, ~~\langle (S_1^z)^2(S_r^z)^2 \rangle = 0.03125.
\end{eqnarray}
These correlation functions are nontrivial, and it seems that they appear for 
the first time. One may note that these properties are quite different from
those of Ising model\cite{baxter,ising}.

\subsection{Antiferromagnetic $t-J_z-V$ model at half-filling}

Consider the simplified version of $H_{tJV}$, described by the 
Hamiltonian (18) with $J=0$, which we call the $t-J_z-V$ model. Below we will 
show that the ground states of this simplied model can be constructed using
MP ground state approach at half-filling. As before, we define an $2 \times 2$
matrix 
\begin{eqnarray}
g_{j}''' = \left( \begin{array}{cc}
         |\uparrow\rangle + |\downarrow\rangle & |\uparrow\rangle - |\downarrow\rangle  \\
|\downarrow\rangle - |\uparrow\rangle     & 
- |\downarrow\rangle - |\uparrow\rangle 
                \end{array} \right)_j.
\end{eqnarray}
We propose an ansatz
\begin{eqnarray}
|\Phi_3\rangle = Tr g_1''' \otimes g_2''' \otimes \cdots  \otimes g_L'''.
\end{eqnarray}
Pursuing the same procedure as in Sec.II, one can find that $|\Phi_3\rangle $
is really the global ground state, with eigenvalue $(V-\frac{J_z}{4})L$, of the
$t-J_z-V$ model, subject to the following constraint
\begin{eqnarray}
J_z >0, ~~t>0, ~~V \leq -t + \frac{J_z}{4}.
\end{eqnarray}
Similar to the arguments in Ref.\cite{klu}, it can be shown that the global ground state is unique.
The expectation values in the ground state are obtained
\begin{eqnarray}
\langle S_j^z \rangle =0, ~~ \langle S_{tot}^z \rangle = \langle \sum_{j=1}^L S_{j}^z \rangle =0, ~~ \langle S_j^{\pm} \rangle = 0,~~
\langle n_j \rangle =1,~~ \langle (S_j^z)^2 \rangle =\frac14.
\end{eqnarray}
The global ground state is thus antiferromagnetic. The spin-spin correlation
function is found to be
\begin{eqnarray}
\langle S_1^zS_r^z \rangle = - \frac14 (-1)^r, ~~(r \geq 2)
\end{eqnarray}
which is nothing but the result of Ising model\cite{baxter,ising}. This fact
shows that the antiferromagnetic $t-J_z-V$ model at half-filling has the same 
properties as the Ising model, implying that the model at half-filling can be 
mapped to the Ising model. Actually, this is true, because the hopping term, 
at half-filling, plays no role in the restricted Hilbert space without 
double-occupied sites. The other two-point correlation functions are
\begin{eqnarray}
\langle S_1^+S_r^- \rangle = \langle S_1^-S_r^+ \rangle = 0, ~~\langle n_1n_r
\rangle = 1, ~~\langle S_1^zn_r \rangle = 0,  ~~\langle (S_1^z)^2(S_r^z)^2 \rangle = 0.03125.
\end{eqnarray}
One may observe that the behaviors of correlators at half-filling are quite
different from those at 2/3 filling, as discussed in last subsection.

\section{Summary and Conclusion}

We have studied the exact ground-state properties of a few 1D quantum systems
including a frustrated spin-1/2 Heisenberg ladder, a ferromagnetic t-J-V model,
an antiferromagnetic $J_z-V$ model at 2/3 filling and an antiferromagnetic 
$t-J_z-V$ model at half-filling, by using the matrix product ground state 
approach. The correlation functions were obtained in the ground states.

It is shown that the ground state of the frustrated spin-1/2 Heisenberg 
ladder with restricted parameters is magnetic ordered and is degenerate. The
excitation from the ground state may be gapless. The correlation
functions are uniform in spatial space, implying that the system is critical
in the ground state in the sense that the correlation length is infinite. 

The exact ground state of the ferromagnetic t-J-V model with proper parameters
was constructed at half-filling. It is found that the ground state has magnetic long-range 
order, and is also degenerate. A phase transition is found in the system. 
The spin rotational symmetry is  
spontaneously broken in the state. The correlation function is uniform, implies
that the system is also critical in the ground state.

The global ground state of the antiferromagnetic $J_z-V$ model with $V=J_z/4$
is found to be antiferromagnetic and unique at 2/3 filling. The spin-spin
correlation functions is fluctuating, non-vanishing eventually, 
with increasing the spatial distances, suggesting that the ground state has
antiferromagnetic long-range order. The spin-charge correlation function is
also fluctuating, implying that the spin and charge degrees of freedom are
highly correlated in the state. The spin and charge are not seperated. The
behaviors are quite different from those of Ising model.

The studies on the antiferromagnetic $t-J_z-V$ model with proper parameters 
at half-filling show that the system is very similar to the Ising model, as
the spin-spin correlation function is exactly the same. The ground state is
unique and has antiferromagnetic long-range order.

\acknowledgments
The author is indebted to Dr. A. Kl\"umper and Dr. A. Schadschneider for 
drawing his attention to this field, helpful 
discussions and advice. He is also grateful to Prof. J. Zittartz and ITP of
Universit\"at zu K\"oln for warm hospitality. This work is supported by the 
Alexander von Humboldt Stiftung.

\end{document}